%% file: main.tex
\pdfoutput=1
\documentclass[11pt]{article}

\usepackage[T1]{fontenc}
\usepackage[utf8]{inputenc}
\usepackage[margin=1in]{geometry}
\usepackage{amsmath,amssymb}
\usepackage{graphicx}
\usepackage{subcaption}
\usepackage{booktabs}
\usepackage{float}
\usepackage[numbers,sort&compress]{natbib}
\usepackage{listings}
\usepackage[hidelinks]{hyperref}

\graphicspath{{figures/}}
\newcommand{\keywords}[1]{\par\noindent\textbf{Keywords:} #1}

\title{Stone-in-Waiting: A Cloud-Based Accelerator for the Quantum Approximate Optimization Algorithm}
\author{%
    Shuai Zeng \\
    Chongqing University of Posts and Telecommunications, Chongqing 400065, China \\
    \texttt{zengshuai@cqupt.edu.cn}
}
\date{}

\begin{document}

\maketitle

\begin{abstract}
The Quantum Approximate Optimization Algorithm (QAOA) and its advanced variant, the Quantum Alternating Operator Ansatz (QAOA), are major research topics in the current era of Noisy Intermediate-Scale Quantum (NISQ) computing. However, the problem of initializing their parameters remains unresolved.
Motivated by the combinatorial optimization task in the 6th MindSpore Quantum Computing Hackathon (2024), this paper proposes Stone-in-Waiting, a cloud-based accelerator for obtaining high-quality initial parameters for QAOA.
Internally, the accelerator builds on state-of-the-art theories and methods for parameter determination and integrates four self-developed algorithms for QAOA parameter initialization, mainly based on Bayesian methods, nearest-neighbor methods, and metric learning. Compared with the Baseline Algorithm, the generated parameters improve the score by 40.19\%.
Externally, the accelerator offers both a web interface and an API, providing flexible and convenient access for users to test and develop related experiments and applications.
This paper presents the design principles and methods of Stone-in-Waiting, demonstrates its functional characteristics, compares the strengths and weaknesses of the four proposed algorithms, and validates the overall system performance through experiments.

\end{abstract}

\input{body}

\end{document}

%% file: body.tex
\section{Introduction}
Optimization methods, as the theoretical and algorithmic foundation of big data and machine learning, have long been a central topic in data science. Humanity is currently in the era of Noisy Intermediate-Scale Quantum (NISQ) computing, and the Quantum Approximate Optimization Algorithm (QAOA) is regarded as one of the most promising quantum optimization algorithms capable of outperforming classical methods\cite{qaoa2014,qaoa2019}. The basic idea of this method is to exploit the properties of quantum adiabatic evolution by controlling or simulating an appropriate quantum physical evolution process to perform optimization. More specifically, the system is first prepared in the ground state of a simple Hamiltonian. Parameterized quantum gates generated from the simple Hamiltonian and from the problem Hamiltonian are then applied alternately for multiple rounds, where the number of rounds is referred to as the circuit depth. In this way, the quantum state is gradually driven toward the ground state of the problem Hamiltonian, thereby solving the optimization problem. One of the main challenges of QAOA is how to determine appropriate parameters for the parameterized quantum gates at each layer accurately and efficiently.\\
\indent
In the original approach, parameter optimization begins with random initial parameters and then relies on a classical optimizer, such as gradient descent, for iterative refinement. However, random initialization has several drawbacks. On the one hand, random starting points are usually far from the optimum, so performance is difficult to guarantee. On the other hand, the optimization process can easily fall into barren plateaus or become trapped in unsatisfactory local optima\cite{BarrenplateausNature2018,galda2023similaritybasedparametertransferabilityquantum}, leading to slow execution or degraded final parameter quality, and thus reducing the overall efficiency and performance of the algorithm. Consequently, non-random methods for parameter initialization have attracted increasing attention in recent years. Studies such as \cite{Harrigan_2021,Wang_2018,Wurtz_2021,Marwaha_2021} have explored the values and distributions of optimal parameters under different circuit depths and analyzed the associated mathematical properties and algorithmic performance. References \cite{D-regular_graphs2022_1,D-regular_graphs2022_2} establish quantitative relationships between parameter values and algorithmic performance for unweighted regular graphs (in the context of this paper, graphs and Hamiltonians are interchangeable and are sometimes used interchangeably in the text) and the Sherrington-Kirkpatrick graph model. They propose an iterative algorithm for directly computing initial parameters for a given regular graph and theoretically prove that when the quantum circuit depth exceeds 11 layers, QAOA can outperform all currently known classical algorithms. Reference \cite{boulebnane2021predictingparametersquantumapproximate} studies parameter determination for unweighted graphs under the Erdos-Renyi model and more general distributional settings, presents related methods, and compares algorithmic performance under different parameter choices. Reference \cite{galda2023similaritybasedparametertransferabilityquantum} experimentally demonstrates the feasibility of parameter transfer in the unweighted-graph setting: for similar graphs, the corresponding parameter values are also very close, so the optimized parameters of one graph can be transferred to a class of similar graphs. Reference \cite{Shaydulin_2023} formalizes this transferability and extends it to weighted graphs by proposing a parameter-scaling algorithm that converts and reuses optimized parameters obtained from unweighted graphs. The most recent work\cite{Sureshbabu_2024} further improves the parameter-scaling algorithm for weighted graphs and extends it to higher-order hypergraph applications, enabling efficient generation of initial parameters for higher-order QAOA. This method is currently the most advanced and effective approach for generating initial QAOA parameters, and it is also the Baseline Algorithm for the combinatorial optimization task of the 6th MindSpore Quantum Computing Hackathon. Taken together, these theoretical and experimental studies point to the same direction: whether for unweighted or weighted graphs, and across different graph types or edge-weight distributions, similar graphs tend to have similar optimized parameters. In other words, optimized parameters known for one graph can be transferred, by appropriate means, to more unknown graphs that are similar to it.\\
\indent
Motivated by this idea, this paper proposes and designs a cloud-based accelerator for QAOA. Building on the solution to the combinatorial optimization task of the 6th MindSpore Quantum Computing Hackathon, the accelerator further provides a web interface and a Python-based API, enabling researchers and developers to obtain QAOA initialization parameters conveniently and flexibly, and thereby share efficient parameters derived from identical or similar graphs across the community. Internally, on top of the latest techniques for generating initial parameters for higher-order QAOA described above, the system designs and implements four parameter-generation sub-algorithms. By integrating and selecting among these sub-algorithms, the system achieves high-quality QAOA initial parameters with performance significantly above the Baseline Algorithm.\\
\indent
Its functionality and design details further include:
\\ \indent(1) Continuous parameter search
\\ \indent Since denser coverage of similar graphs leads to better parameter transfer, the cloud server continuously searches for optimized parameters for various graph models during idle periods, accumulating increasingly accurate and efficient parameters over time for future use and storage;
\\ \indent(2) Chained genetic optimization
\\ \indent Because parameters can be transferred among similar graphs, the search process can start from some typical graphs and then expand gradually to nearby graphs through genetic mutations, rather than optimizing every graph independently from scratch. This process combines the ideas of neutron chain reactions and biological inheritance, further improving the efficiency of parameter discovery and creating a scale effect;
\\ \indent(3) Reverse parameter updating
\\ \indent After obtaining initial parameters from the accelerator, users, namely researchers and developers, may further optimize them using their own computational resources, such as private quantum computers or edge devices, to obtain more precise parameters. Once their effectiveness is verified, the accelerator allows users to upload these refined parameter data through both the web interface and the API, so that the accelerator can share and output increasingly accurate parameters;
\\ \indent(4) Distributed computing
\\ \indent Multiple accelerators can be deployed in parallel. Different search tasks can be assigned across them, and the resulting parameters can be shared to form a parameter network at the industry or even global scale;
\\ \indent(5) Decoupling of internal and external modules
\\ \indent The internal algorithms and external services are designed as decoupled modules. When new parameter-generation techniques appear, the internal algorithms can be upgraded at any time to improve service quality, while changes to external service items or access modes do not affect internal operation.\\
\indent
As naming research projects after figures from ancient Chinese mythology or classical allusions has become common in China, and because the cloud-based accelerator proposed in this paper is only a modest design, we borrow the name "Stone-in-Waiting," a minor divine figure from the Chinese classical novel \textit{Dream of the Red Chamber}. The name reflects the idea that just as Stone-in-Waiting is devoted to watering and nurturing the Crimson Pearl Flower, the proposed accelerator is devoted to providing high-quality initialization parameters for QAOA so as to accelerate and support quantum computing. The domain name of the accelerator, stone-in-waiting.sbs, is also derived from the English translation "Divine Luminescent Stone-in-Waiting."\\
\indent
The remainder of this paper is organized as follows. Section 2 introduces how to use the proposed accelerator, including the web interface and API examples. Section 3 presents the design principles, methods, and implementation details of the accelerator. Section 4 provides experimental comparisons to justify the design choices and validate the effectiveness of the accelerator. Section 5 concludes the paper and discusses future work.

%%%%%%%%%%%%%%%%%%%%%%%%%%%%%%%%%%%%%%%%%%%%%%%%%%%%%%%%%%%%% 

%%%%%%%%%%%%%%%%%%%%%%%%%%%%%%%%%%%%%%%%%%%%%%%%%%%%%%%%%%%%% 

\section{Usage}
\subsection{Web Access}
The web interface of Stone-in-Waiting is simple and easy to use. By visiting stone-in-waiting.sbs, users can directly access the functional interface. The current web interface provides three basic functions.\\
\indent (1) Parameter Query\\
\indent Users can submit graph data in the "Graph Data" text box under "Parameter Query", select the quantum circuit depth from the drop-down menu below, and click the "Query" button to retrieve the accelerator-optimized initial parameters corresponding to the input graph data (the specific generation algorithms are described later).
The description of "Graph Data" follows the format specified in the 6th MindSpore Quantum Computing Hackathon and is packaged as a Python Dict. The Dict contains a key named 'J' and a key named 'c', where 'J' corresponds to the edge set of the graph described by vertex pairs, and 'c' corresponds to the associated edge weights. For example, the graph data \{'J': [[5, 9], [1, 2], [8, 11]], 'c': [5, 6, 7]\} represents a graph in which the edge between nodes 5 and 9 has weight 5, the edge between nodes 1 and 2 has weight 6, and the edge between nodes 8 and 11 has weight 7.
If the parameter query succeeds, the system returns the status "success" together with the parameter values, where the parameter values are packaged as a Python List. For a graph with circuit depth 4, the returned List contains 8 floating-point numbers. The gamma parameters at different depths are the 0th, 2nd, 4th, and 6th elements of the List, while the beta parameters are the 1st, 3rd, 5th, and 7th elements. Similarly, for a graph with circuit depth 8, the returned List contains 16 floating-point numbers. The gamma parameters and beta parameters at each depth are the even-indexed and odd-indexed elements of the List, respectively.\\
\indent (2) Parameter Submission\\
\indent If users obtain parameter values that are better than the initial parameters currently provided by the accelerator, they can submit them through the "Parameter Submission" function to update the accelerator data. Users enter graph data (in Dict format) in the "Graph Data" text box under "Parameter Submission", enter the new parameter values (in List format) in the "Parameters" text box, select the circuit depth, and then click the "Submit" button to complete the submission.
The system evaluates whether the submitted parameter values are indeed better than the existing ones. If the submitted parameters achieve a higher score than the existing parameters, the system returns the status "success" together with a comparison between the old and new parameter scores. Otherwise, it returns the status "fail" and indicates that the submitted parameters are of lower quality than the existing ones. Once a new parameter submission succeeds, the accelerator data are updated, and subsequent queries are computed and returned using the new parameters, thereby realizing the reverse parameter updating function introduced earlier.\\
\indent (3) Parameter Comparison\\
\indent Sometimes users may need a quantitative evaluation of the performance of certain parameter values. In this case, they can use the "Parameter Comparison" function on the web interface. Users enter graph data (in Dict format) in the "Graph Data" text box under "Parameter Comparison", enter the parameter values to be compared (in List format) in the "Parameters" text box, select the circuit depth, and then click the "Compare" button.
If the operation succeeds, the system returns the status "success" together with three scores: "Current Best Parameter Score", "Uploaded Parameter Score", and "Random Parameter Score". Here, "Current Best Parameter Score" is the score of the best initial parameters currently known to the accelerator, "Uploaded Parameter Score" is the score of the parameters submitted by the user, and "Random Parameter Score" is the score of a randomly generated set of parameters. These three scores allow users to make a rough assessment of the performance of the submitted parameters.

\subsection{API Programming Interface}
To further facilitate researchers and developers in implementing QAOA, and potentially to help this approach become a standard paradigm for QAOA quantum algorithms in the future, Stone-in-Waiting provides a more direct and efficient API programming interface in addition to the web interface. When used properly, this API can substantially improve the efficiency of QAOA parameter optimization.
If the process of QAOA parameter optimization is compared to climbing a mountain, then making full use of the Stone-in-Waiting API is equivalent to taking a cable car to a platform near the summit, greatly reducing the effort and time required for the final climb. Moreover, if this method can become an industry standard and be widely adopted after quantum computers become practical,
it could greatly save computational resources and improve the speed and accuracy of quantum computing.
\\ \indent This API is developed in Python and is very easy to install. For Python 3.6 and above, it can be installed using the following pip command.
\begin{verbatim}
                    pip install stone_in_waiting 
\end{verbatim}
Corresponding to the web interface, the API provides the three interface functions shown in Listing 1.
The function API\_query\_parameter is used for parameter queries. Its two input arguments are graph data and the quantum circuit depth. The graph data format is the same as that used by the web interface and follows the format specified in the 6th MindSpore Quantum Computing Hackathon.
Its two return values are the query status and the parameter values. A successful query returns the status "success", a failed query returns "fail", and an error returns "error". The parameter values are returned in the form of a List.
The function API\_submit\_parameter is used for parameter submission. Its three input arguments are the graph data, the submitted parameters, and the quantum circuit depth.
Its return values are the submission status and the score. The status "success" indicates that the submission succeeded and the new parameters have taken effect in the system. The status "fail" indicates that the score of the submitted parameters is less than or equal to the score of the parameters currently generated by the system, while the status "error" indicates that an error occurred.
The score is returned as a Dict of the form \{'max\_score':max\_score, 'user\_score':user\_score\}, where the key 'max\_score' corresponds to the score of the system-generated parameters before the new submission, and the key 'user\_score' corresponds to the score of the new parameters submitted by the user.
The function API\_compare\_parameter is used for parameter comparison. Its three input arguments are the graph data, the submitted parameters, and the quantum circuit depth.
Its return values are the comparison status and the score. A successful comparison returns the status "success", a failed comparison returns "fail", and an error returns "error". The score is returned as a Dict of the form \{'max\_score':max\_score, 'user\_score':user\_score, 'random\_score':random\_score\},
where the key 'max\_score' corresponds to the score of the parameters currently generated by the system, the key 'user\_score' corresponds to the score of the parameters submitted by the user, and the key 'random\_score' corresponds to the score of a randomly generated parameter set.
\begin{lstlisting}[language=Python, caption=Definition of API Interface Functions]
    import requests
    url = 'http://stone-in-waiting.sbs/api'

    def API_query_parameter(graph_data, qc_depth):
        data_dict = {'api_name': 'query_parameter', 'graph_data':graph_data, 'qc_depth':qc_depth}
        response = requests.post(url, json=data_dict).json()
        return response['status'], response['parameter']
    
    def API_submit_parameter(graph_data, user_parameter, qc_depth):
        data_dict = {'api_name': 'submit_parameter', 'graph_data':graph_data, 'user_parameter':user_parameter,'qc_depth':qc_depth}
        response = requests.post(url, json=data_dict).json()
        return response['status'], response['score_dict']
    
    def API_compare_parameter(graph_data, user_parameter, qc_depth):
        data_dict = {'api_name': 'compare_parameter', 'graph_data':graph_data, 'user_parameter':user_parameter,'qc_depth':qc_depth}
        response = requests.post(url, json=data_dict).json()
        return response['status'], response['score_dict']
\end{lstlisting}

\begin{figure} [H]
    \centering 
    \includegraphics[width=1\textwidth]{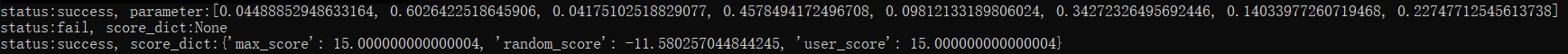}
    \caption{Execution Result of the API Interface Example}
    \label{fig:apiclient}
\end{figure}

This API is very easy to use. After installation, it can be imported into a program with the following code.
\begin{verbatim}
                    from stone_in_waiting import *
\end{verbatim}
Listing 2 provides a complete example of how to use the API interface functions. The example code demonstrates the parameter query, parameter submission, and parameter comparison functions of the API, and the execution result is shown in Figure~\ref{fig:apiclient}.\\ \indent

\begin{lstlisting}[language=Python, caption=Example of Using the API Interface Functions]
    from stone_in_waiting import *

    if __name__ == '__main__':
        graph_data = {'J': [[5, 9], [1, 2], [8, 11]], 'c': [5, 5, 5]}
        qc_depth = 4

        status, parameter = API_query_parameter(graph_data, qc_depth)
        print(f'status:{status}, parameter:{parameter}')
        
        user_parameter = parameter
        status, score_dict = API_submit_parameter(graph_data, user_parameter, qc_depth)
        print(f'status:{status}, score_dict:{score_dict}')
        
        status, score_dict = API_compare_parameter(graph_data, user_parameter, qc_depth)
        print(f'status:{status}, score_dict:{score_dict}')
\end{lstlisting}

Due to the author's limited time and capacity, the hurried development process, and the weak server hardware, the current Stone-in-Waiting version 0.0.1v still has a rudimentary interface, relatively slow computation, and a rather limited service scope. The current work is mainly targeted at the problem of the 6th MindSpore Quantum Computing Hackathon and currently supports only the initial parameters of graphs with 12 nodes and quantum circuit depths of 4 and 8.

%%%%%%%%%%%%%%%%%%%%%%%%%%%%%%%%%%%%%%%%%%%%%%%%%%%%%%%%%%%%% 

\section{System Design}
\subsection{Overview}
\begin{figure} 
\centering 
\includegraphics[width=1\textwidth]{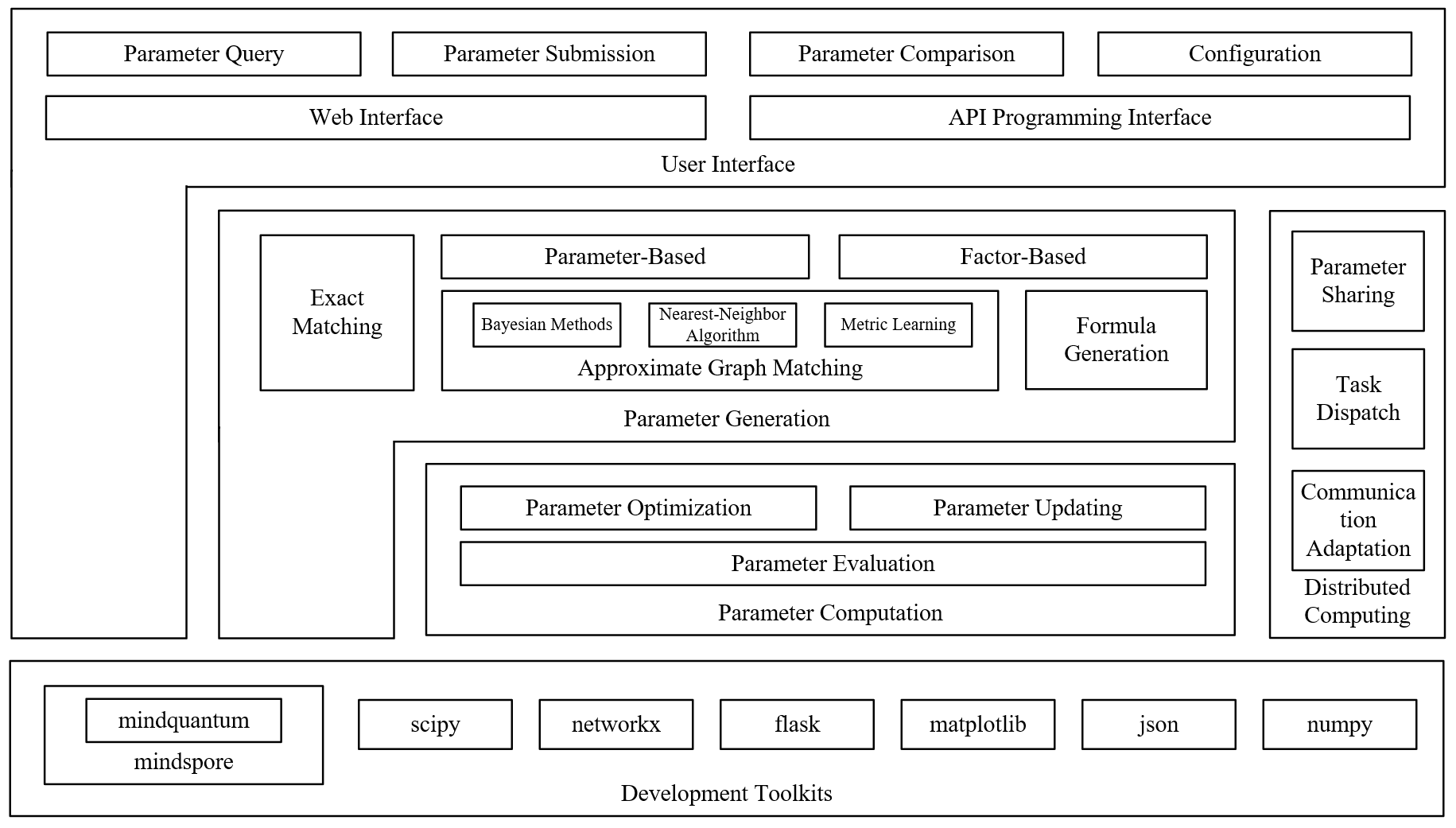}
\caption{Overall Architecture of Stone-in-Waiting}
\label{fig:ztmkt}
\end{figure}

The overall functional architecture of Stone-in-Waiting is shown in Figure~\ref{fig:ztmkt}.
First, the system uses Python as its programming language.
The quantum-computing components related to QAOA are mainly developed based on MindSpore's mindquantum toolkit,
graph-related operations are mainly developed based on the networkx toolkit, and machine-learning-related operations are mainly developed based on the scipy toolkit.
The web interface, API programming interface, and distributed computing functions are developed based on the flask toolkit,
while basic computation, graphical display, and file reading and writing are developed based on numpy, matplotlib, and json.
Built on top of these foundational toolkits is the parameter computation module, whose role is to compute optimized parameters for various graphs for subsequent use.
Its main basic functions are to return the score of a given graph and parameter set, and on top of that, to perform parameter optimization and parameter updating.
Parameter optimization itself is further divided into multiple specific execution algorithms. In practice, alternating among different algorithms over multiple rounds can continuously improve parameter quality.
On top of the parameter computation module is the parameter generation module. Stone-in-Waiting contains four independent parameter-generation methods, and this part is the core that enables the accelerator to produce high-quality initial parameters. It will be introduced in detail later.
In addition, Stone-in-Waiting also includes a distributed computing module, so that when multiple systems can be deployed, parameter optimization and generation can be carried out in parallel.
Finally, the top layer of the system is the user interface module. Through both the web interface and the API programming interface, this module provides researchers and developers with services such as parameter query, parameter submission, parameter comparison, and system configuration.
\\ \indent 
At present, Stone-in-Waiting is still in the early stage of development, and version 0.0.1v has not yet fully implemented all of the designed features. The existing functions are mainly concentrated in the parameter computation and parameter generation parts directly related to the problem of the 6th MindSpore Quantum Computing Hackathon, while the user interface remains relatively simple and the distributed computing part is still under development.
\subsection{Major Features and Design Details}
\subsubsection{Parameter Computation Module}
\begin{figure} 
    \centering 
    \includegraphics[width=0.7\textwidth]{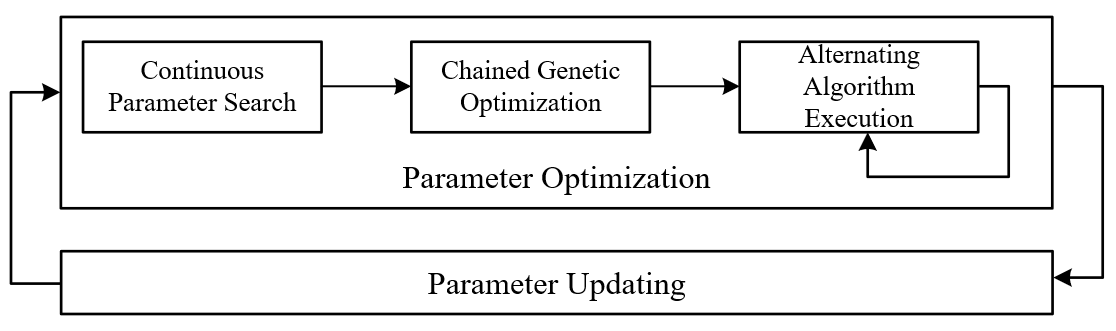}
    \caption{Parameter Computation Process}
    \label{fig:csjsgc}
\end{figure}

The parameter computation module consists of three submodules: parameter evaluation, parameter optimization, and parameter updating. Among them, parameter evaluation is relatively simple. Its purpose is to score a given set of parameters by directly constructing the quantum circuit with the development toolkit, inserting the parameters, executing the circuit, and finally performing measurement. Built on this basic capability, the parameter optimization and parameter updating submodules employ three specific techniques.
\\ \indent (1) Continuous Parameter Search
\\ \indent Since both theory and practice have shown that similar graphs tend to have similar parameters, the more known graph parameters are available, the easier it is to find the graph most similar to the target Hamiltonian. Therefore, the more known graph parameters are accumulated during idle periods, the more accurate the generated parameters for unknown graphs will be on average, and the more significant the acceleration effect on quantum computing will become.
Stone-in-Waiting therefore uses idle server computing resources to carry out continuous routine parameter optimization, continuously exploring parameters for various typical graphs and expanding the graph database of computed parameter values;
\\ \indent (2) Chained Genetic Optimization
\\ \indent To further improve search efficiency, this process of parameter exploration and updating adopts a chained genetic algorithm at the microscopic level.
Starting from some typical graphs, the algorithm generates similar child graphs through genetic mutation. These child graphs inherit the parent graph's parameters as initial values, which are then fine-tuned. Since similar graphs have similar parameters, the computational cost of fine-tuning is small, which can significantly improve search speed.
Moreover, this process can expand recursively in a chain-like manner: child graphs can further mutate and split into more similar child graphs. This process resembles a neutron chain reaction and can rapidly accumulate and expand the coverage of the graph database with computed parameters;
\\ \indent (3) Alternating Algorithm Execution
\\ \indent Empirical tests show that optimization methods based on quantum gradients and methods that do not use quantum gradients, together with different specific optimization algorithms (such as COBYLA and BFGS) and different learning rates, can repeatedly alternate while inheriting parameters from one another during the parameter search process. This makes it possible to squeeze out additional score improvements and further refine accurate parameter values.
The operation of the parameter optimization and parameter updating submodules is shown in Figure~\ref{fig:csjsgc}. The continuous parameter search function formulates the overall plan for parameter optimization, the chained genetic optimization function realizes the concrete computational path of parameter optimization, and alternating algorithm execution guarantees the quality of parameter optimization. After the parameters are obtained,
the process enters the parameter updating module. In essence, the parameter updating module updates the graph database of computed parameters. More specifically, it creates new data or replaces possibly duplicated data and merges them into the existing database. The processes of parameter optimization and parameter updating alternate repeatedly under the action of continuous parameter search.
\subsubsection{Parameter Generation Module}
\begin{figure}[tbp]
    \centering
    \includegraphics[width=0.78\textwidth]{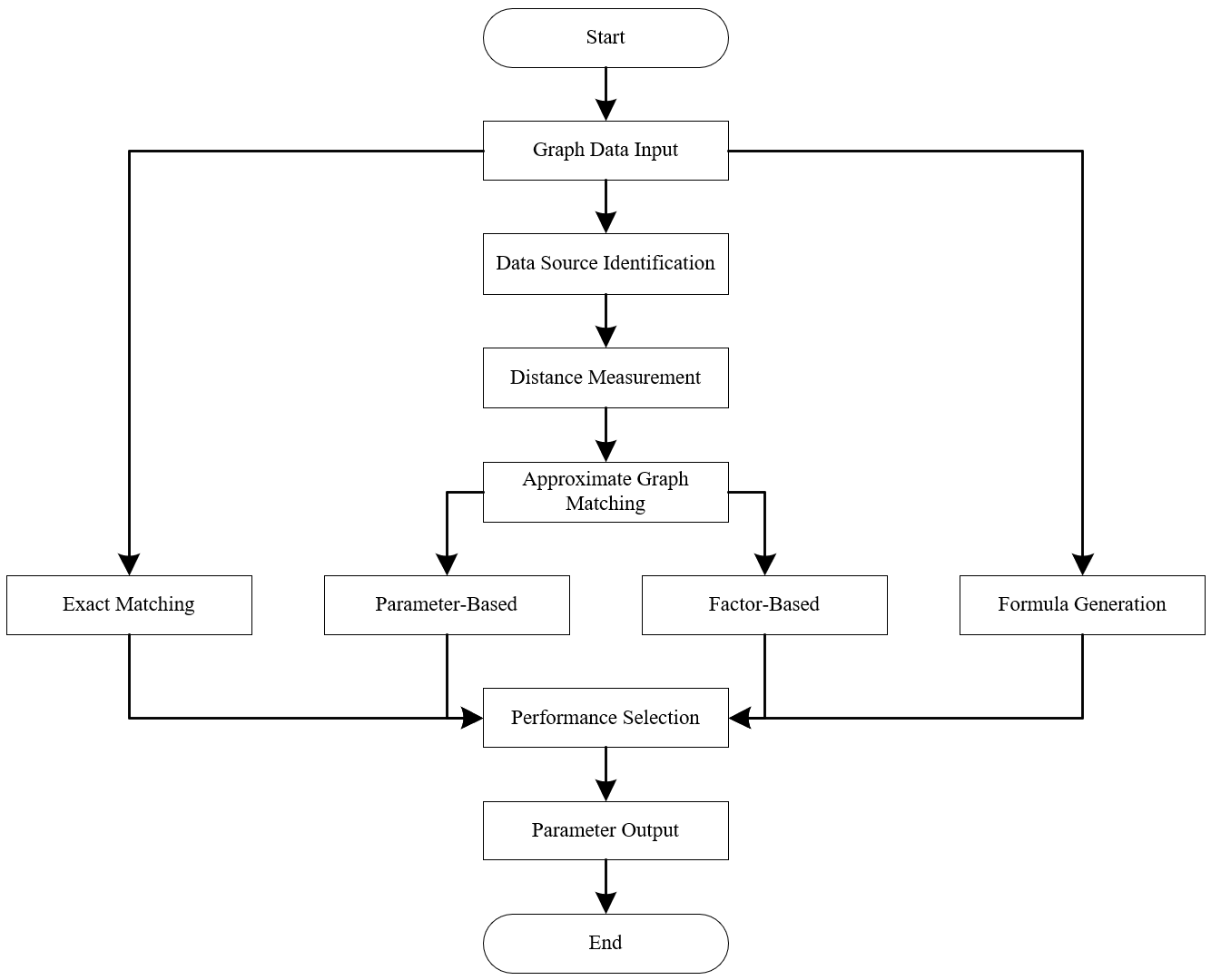}
    \caption{Overall Flowchart of the Parameter Generation Module (Including Parallel Processes)}
    \label{fig:csscmklct}
\end{figure}

The parameter generation module is the key reason Stone-in-Waiting can produce high-quality initial parameters. In practice, this module consists of four independent parameter-generation sub-algorithms, referred to as Exact Matching, Parameter-based Approximate Graph Matching, Factor-based Approximate Graph Matching, and Formula Generation.
First, these four sub-algorithms independently generate parameters. The quality of the generated parameters is then evaluated, and the best-performing set is selected as the module output. The specific workflow of the parameter generation module is shown in Figure~\ref{fig:csscmklct}. Since the four parameter-generation sub-algorithms can operate independently in logic, the flowchart includes parallel execution relationships for clarity.
As can be seen from the flowchart, after the graph data are read in, the graph data are duplicated and sent to three parallel branches. Among them, the Exact Matching and Formula Generation algorithms can directly obtain parameters from the graph data, whereas the Parameter-based Approximate Graph Matching and Factor-based Approximate Graph Matching algorithms first require data source identification, distance measurement, and approximate graph matching.
Based on the matching results, the program then enters the parameter-based and factor-based algorithmic branches in parallel, and these branches output parameters separately. Finally, the parameters produced by all four sub-algorithms are collected at the performance-selection node, and the best-performing parameters are chosen as the final output of the module.
\\ \indent The principles of these four proposed sub-algorithms are introduced below in detail. 
\\\indent (1) Exact Matching
\\ \indent Exact Matching is the simplest and most direct of the four algorithms. It directly compares the input graph data with graph data for which parameters have already been computed. If the input graph exists in the database of graphs with computed parameters, the corresponding parameters are returned. Clearly, the advantage of this method is that the process is simple, the complexity is low, and the quality of the output parameters is extremely high.
However, the probability of finding an exact match is low, so the algorithm lacks generalization ability, and its performance depends to a large extent on the scale of the graph database with computed parameters. The continuous parameter search function introduced earlier in the parameter computation module provides useful support for this method. At present, this method is mainly applied to the local dataset of the 6th MindSpore Quantum Computing Hackathon.
\\ \indent (2) Parameter-based Approximate Graph Matching
\\ \indent The basic motivation of the proposed Parameter-based Approximate Graph Matching algorithm is the principle, summarized earlier from existing frontier literature, that similar graphs have similar parameters. Its specific implementation is an organic combination and integrated application of multiple machine learning algorithms to the problem of the 6th MindSpore Quantum Computing Hackathon.
\\ \indent (a) Data Source Identification
\\ \indent The algorithm first needs to identify the "source" of the input graph data, namely to determine the graph order, edge-weight distribution, edge-generation probability, number of nodes, number of edges, and related information. Among these, the graph order, number of nodes, and number of edges can be obtained directly from the graph data, whereas the edge-weight distribution and edge-generation probability can be obtained through Bayes' formula and maximum likelihood estimation.
\\ \indent Let $\mathcal{F}$ be the set of edge-weight distribution functions, and let $f\in \mathcal{F}$ be a possible edge-weight distribution function. Given the dataset $W$ of edge weights in the graph data, the likelihood that $W$ comes from the weight-distribution function $f$ can be written as $P(f|W)$. According to Bayes' formula, we have Eq. (\ref{eq:bys}), and its maximum likelihood estimate is given by Eq. (\ref{eq:jdsrgj}).
Assume that all distribution functions in $\mathcal{F}$ are equally likely and that all edge weights are independent and identically distributed. That is, for all $f\in \mathcal{F}$, $P(f)$ is unchanged, and $P(W|f)$ can be written as a product over all edges. Since $P(W)$ is independent of the value of $f$, let the edge weights be $w_1$,$w_2$,...,$w_{|W|}$ respectively, where $|W|$ denotes the total number of edges in the graph.
Then Eq. (\ref{eq:jdsrgj2}) follows. In this way, by substituting the possible distribution functions and the observed weight data, one can determine which distribution the edge weights are most likely to come from. When the number of edges is large, direct multiplication of probabilities can easily overflow, so the log form is typically used, and the problem is solved with the log-likelihood formula in Eq. (\ref{eq:ds_jdsrgj}). Finally, the edge-generation probability can also be obtained in a similar way. At this point, the data source identification process is complete,
and it outputs the graph order, edge-weight distribution, edge-generation probability, number of nodes, number of edges, and related information as the input for the next step of distance measurement.
\begin{equation}
    P(f|W)=P(W|f)P(f)/P(W) \label{eq:bys}  
\end{equation}
\begin{equation}
    \mathop {\arg \;\max }\limits_{f \in \mathcal{F}} P(f|W) =\mathop {\arg \;\max }\limits_{f \in \mathcal{F}} P(W|f)P(f)/P(W)  \label{eq:jdsrgj} 
\end{equation}
\begin{equation}
    \mathop {\arg \;\max }\limits_{f \in \mathcal{F}} P(f|W) =\mathop {\arg \;\max }\limits_{f \in \mathcal{F}} \prod_{i=1}^{|W|} P(w_i|f)  \label{eq:jdsrgj2} 
\end{equation}
\begin{equation}
    \mathop {\arg \;\max }\limits_{f \in \mathcal{F}} P(f|W) =\mathop {\arg \;\max }\limits_{f \in \mathcal{F}} \sum_{i=1}^{|W|}  \log P(w_i|f)  \label{eq:ds_jdsrgj} 
\end{equation}
\\ \indent (b) Distance Measurement
\\ \indent The graph order, edge-weight distribution, edge-generation probability, number of nodes, number of edges, and related information obtained from data source identification can be regarded as the attributes of a graph. The set of all possible attribute values then forms a graph space $\mathcal{G}$. The attributes of a graph correspond to a coordinate in this graph space.
From a qualitative perspective, the distance between two graphs can therefore be defined by the differences in these attribute values. Quantitatively, when the attributes play comparable roles, the Euclidean distance between attribute values can be used as the distance measure. Let $g_i, g_j \in \mathcal{G}$ be any two graphs in graph space, where $g_i$ and $g_j$ are the coordinate vectors of the two graphs.
Let $dist_2(g_i, g_j)$ denote the Euclidean distance between the two graphs. Then the distance formula can be written as Eq. (\ref{eq:osjl}).
\begin{equation}
    dist_2(g_i, g_j)= \sqrt{\left \| g_i-g_j \right \|_{2}^{2}}= \sqrt{(g_i-g_j)^{T}(g_i-g_j) }   \label{eq:osjl}  
\end{equation}
\indent However, defining graph distance in this way presents two problems. First, the effects of different graph attributes are not necessarily balanced. For example, the influence of the number of edges on the parameters may be greater than that of the graph order. Second, these attributes are not necessarily independent and may be correlated. For example, the edge-generation probability may be correlated with the number of edges.
Therefore, when the attribute space is relatively complex, it is more appropriate to use the Mahalanobis distance to define the distance between two graphs. Let $M$ be a symmetric positive semidefinite matrix, and let $dist_M(g_i, g_j)$ denote the Mahalanobis distance between the two graphs. Then the distance formula can be written as Eq. (\ref{eq:msjl}).
\begin{equation}
    dist_M(g_i, g_j)= \sqrt{\left \| g_i-g_j \right \|_{M}^{2}}= \sqrt{(g_i-g_j)^{T}M(g_i-g_j) }   \label{eq:msjl}  
\end{equation}
\indent The next question is how to determine an appropriate $M$. Since the goal of the algorithm is to transfer parameter values among approximate graphs, and since the score of a parameter set is the criterion used to measure parameter quality, the score difference obtained by applying the same parameters to different graphs can be treated as the true value of the distance, while the Mahalanobis distance is treated as the predicted value.
By minimizing the difference between the true value and the predicted value, one can determine $M$.
\\ \indent Let the score of the optimal parameters of $g_i$ be $s_i$, and let the score obtained by applying these parameters to $g_j$ be $s_j^i$. Similarly, let the score of the optimal parameters of $g_j$ be $s_j$, and let the score obtained by applying these parameters to $g_i$ be $s_i^j$. Then the score difference between the two graphs
can be represented by Eq. (\ref{eq:zsjl}). The value of $M$ can be determined by Eq. (\ref{eq:msjl_m}).
\begin{equation}
    dist_{true}(g_i,g_j)=|s_i-s_j^i|+|s_j-s_i^j|   \label{eq:zsjl}  
\end{equation}
\begin{equation}
    \mathop {\arg \;\min }_{M}\sum_{g_i,g_j \in \mathcal{G}} \frac{1}{2} (dist_M(g_i, g_j)-dist_{true}(g_i,g_j))^2   \label{eq:msjl_m}  
\end{equation}
\\ \indent Let $n$ be the number of graph-data samples and $m$ the number of graph attributes, that is, the coordinate dimension. Then the complexity of this metric-learning algorithm is $O(nm^2C_n^2)$. When $n$ and $m$ are small, $M$ can be solved directly; when $n$ and $m$ are large, methods such as Monte Carlo approximation can be used to estimate $M$.
\\ \indent Since this metric learning is performed statically and offline only during training, and prediction directly uses the metric matrix $M$, the system is not sensitive to the complexity of this algorithm.
Experiments in the actual implementation show that the number of edges has the greatest influence on the parameters. To simplify the computation, Stone-in-Waiting 0.0.1v currently measures distance by directly comparing the difference in the number of edges under the conditions of identical graph order, edge-weight distribution, edge-generation probability, and number of nodes. In a future, more precise version, the complete algorithm proposed in this design can be used.
\\ \indent (c) Approximate Graph Matching
\\ \indent Using the metric matrix $M$ obtained from the distance-measurement step, one can calculate the distance between two graphs. Then, by applying the nearest-neighbor algorithm, the system can match the input graph data to similar graphs in the database of graphs with computed parameters and output the corresponding parameters.
\\ \indent Let $k$ be the number of neighboring graphs considered by the algorithm, and let $g_0$ be the coordinate of the input graph. Suppose that, through the metric matrix $M$, the $k$ graphs in the database of graphs with computed parameters that are closest to the input graph $g_0$ are $g_1,g_2,...,g_k$.
Let their distances to $g_0$ be represented by $dist_M(g_0, g_i)$ and their known parameters by $p_1,p_2,...,p_k$. Then the parameters of the input graph are determined by Eq. (\ref{eq:jstpp}). At this point, the Parameter-based Approximate Graph Matching algorithm is complete.
\begin{equation}
    p_0 = \frac{\sum_{i=1}^{k} p_i/dist_M(g_0,g_i)}{\sum_{i=1}^{k} 1/dist_M(g_0,g_i)}   \label{eq:jstpp}  
\end{equation}
\indent In the current implementation of Stone-in-Waiting 0.0.1v, for the sake of simplicity, the algorithm mainly uses nearest-neighbor settings with $k=1$ and $k=2$. More settings can be explored in future versions. In general, the basic idea of this algorithm is to find approximate graphs of the input graph in the database of graphs with computed parameters and then combine these known parameters to obtain the parameters of the input graph.
Therefore, the larger the graph database with computed parameters, the more obvious the improvement in algorithm performance, because approximate graphs can be closer to the input graph and parameter prediction can be more accurate. The continuous parameter search, chained genetic optimization, and reverse parameter updating functions in the accelerator are precisely designed to support this principle.
\\ \indent (3) Factor-based Approximate Graph Matching
\\ \indent The Factor-based Approximate Graph Matching algorithm is an improved algorithm based on the most recent literature of this year\cite{Sureshbabu_2024}. Unlike the parameter-based method, this method does not directly use the parameters of approximate graphs, but instead uses the scaling factors of approximate graphs to generate parameters indirectly.
\\ \indent Reference \cite{Sureshbabu_2024} provides a method for directly computing QAOA initial parameters, with the formula given in Eq. (\ref{eq:jxsf}).
In the parameter formula for $\gamma$, the term $\arctan \frac{1}{\sqrt{D-1}}$ is an empirically set factor. However, experiments show that this factor can be further optimized. Similar to the Parameter-based Approximate Graph Matching algorithm, the system first uses search algorithms such as COBYLA and BFGS to find optimized factors for various typical graphs, with the search initialized at $\arctan \frac{1}{\sqrt{D-1}}$, thereby building a database of graphs with computed factors.
Because the graphs in this database are specially optimized with $\arctan \frac{1}{\sqrt{D-1}}$ as the initial value, the final parameter quality is better than that given directly by Eq. (\ref{eq:jxsf}).
\begin{equation}
    \begin{array}{c}
        \gamma=\alpha \times \gamma^{\infty} \times \arctan \frac{1}{\sqrt{D-1}},  \;\ \beta=\beta^{\infty} \label{eq:jxsf}  \\
        (\alpha=\sqrt{\frac{1}{\left|E_{k}\right|} \sum_{\left\{u_{1}, \cdots, u_{k}\right\}}\left(J_{u_{1}, \cdots, u_{k}}^{(k)}\right)^{2}+\cdots+\frac{1}{\left|E_{1}\right|} \sum_{\{u\}}\left(J_{u}^{(1)}\right)^{2}})  
    \end{array}
\end{equation}
\indent Let $k$ be the number of neighboring graphs considered by the algorithm, and let $g_0$ be the coordinate of the input graph. Suppose that, through the metric matrix $M$, the $k$ graphs in the database of graphs with computed factors that are closest to the input graph $g_0$ are $g_1,g_2,...,g_k$.
Let their distances to $g_0$ be represented by $dist_M(g_0, g_i)$ and their known factors by $o_1,o_2,...,o_k$. Then the factor of the input graph is determined by Eq. (\ref{eq:jstpp_yz}). Replacing the factor $\arctan \frac{1}{\sqrt{D-1}}$ in Eq. (\ref{eq:jxsf}) with $o_0$, the parameter $\gamma$ of $g_0$ is determined by Eq. (\ref{eq:jxsf_yz}).
\begin{equation}
    o_0 = \frac{\sum_{i=1}^{k} o_i/dist_M(g_0,g_i)}{\sum_{i=1}^{k} 1/dist_M(g_0,g_i)}   \label{eq:jstpp_yz}  
\end{equation}
\begin{equation}
    \begin{array}{c}
        \gamma=\alpha \times \gamma^{\infty} \times o_0 \label{eq:jxsf_yz}  
    \end{array}
\end{equation}
\indent The Factor-based Approximate Graph Matching algorithm is similar to the Parameter-based Approximate Graph Matching algorithm in that it reuses the results of data source identification, distance measurement, and approximate graph matching.
The difference is that the parameter-based algorithm produces more accurate parameter outputs but has weaker generalization performance, whereas the factor-based algorithm is slightly less accurate. However, because it is essentially a formula-based method, it has better universality.
In regions where the graph database with computed parameters is not dense enough, the factor-based algorithm performs better on average than the parameter-based algorithm.
\\ \indent (4) Formula Generation
\\ \indent This algorithm adjusts some coefficient values in the algorithm of Ref. \cite{Sureshbabu_2024} through experimental and empirical methods, and then directly uses the formula to generate parameters. The output parameter accuracy of this method is lower than that of the first three methods, but it has the widest range of applicability and mainly serves as a fallback when the first three algorithms are not sufficiently effective.
\\ \indent In general, the four proposed sub-algorithms have different strengths, weaknesses, and use cases. The final parameter generation module compares their outputs and selects the highest-quality parameter set for a specific input graph and application scenario.

\subsubsection{User Interface Module}
The basic function of the user interface module is to realize user access through the web interface and to support developers in further development using the API programming interface functions. Specifically, the web interface of Stone-in-Waiting uses flask as the backend framework, while the API programming interface functions are implemented by uploading and registering a custom Python package on PyPI.
At present, this part of the system is still relatively basic. As mentioned earlier, it mainly supports the three services of parameter query, parameter submission, and parameter comparison.
\\ \indent It is worth noting that the parameter submission module actually realizes the reverse parameter updating function described earlier, namely that users can upload parameters that are better than the best parameters
currently output by the accelerator, so as to improve the parameter output performance of the accelerator. The main motivations for designing and implementing this function are as follows:
First, after obtaining initial parameters from the accelerator, users will usually further iterate and refine them using their own computational resources, such as quantum computers or edge devices, in order to reach local or global optima. Therefore, submitting the optimized parameter values
does not impose any additional burden on users. Second, this parameter updating process can be regarded as a form of user-generated data mechanism. Especially under the current circumstance that quantum computing power, or the computing power of classical machines simulating quantum computers, is still quite precious, using programming interfaces to feed back optimized parameters in batches so as to form an algorithmic standard
or even an industry convention would be a highly efficient way to accumulate parameter resources, and could even become a new standard operational paradigm for the environmentally efficient use of quantum computing resources, analogous to the energy recovery system of hybrid vehicles.
\\ \indent In addition, in the implementation of the user interface module, special attention is paid to decoupling the internal and external modules, so that when the core algorithms in the parameter computation and parameter generation modules are updated, the external web interface and API programming interface remain unaffected, and vice versa.

\section{Experimental Comparison}
\subsection{Experimental Environment and Basic Settings}
At present, the deployment environment of Stone-in-Waiting is relatively modest. The host is a shared-CPU cloud server with 500 MB of memory and 10 GB of disk space. The operating system is Debian 11, the software environment is Python 3.9, and the main development toolkits include Mindspore, mindquantum, numpy, scipy, networkx, flask, requests, and matplotlib.
The quantum circuit construction, simulation, and QAOA-related experiments in this paper are implemented based on the MindSpore Quantum framework\cite{Xu_2024_MindSporeQuantum}.
The source code corresponding to the implementation used in this work is available from the following Gitee repository:
\url{https://gitee.com/mindspore/mindquantum/tree/research/hackathon/hackathon2024/qaoa/hackathon_qaoa_new}.
The basic experimental settings are mainly designed for the 6th MindSpore Quantum Computing Hackathon. The number of graph nodes is fixed at 12, and the quantum circuit depth is fixed at 4 and 8. The local dataset consists of the graph data in the \texttt{data} directory provided by the competition, while the online dataset consists of the graph data in \texttt{data/\_hidden} generated after code submission to the competition scoring system.
Since the competition scoring system allows only a few submissions per day, and to facilitate more extensive experiments, 3600 graph instances are randomly generated in the local \texttt{data/\_hidden} directory using the competition-provided \texttt{\_generate\_data.py} file to simulate the online dataset.
The local dataset and the simulated online dataset together constitute the full dataset. During execution, the full dataset is appropriately divided into training, validation, and test subsets according to the needs of specific problems, so as to support various kinds of training and experiments.

\subsection{Experiments on the Design Rationale}
\begin{figure}
    \centering
    \subcaptionbox{Scenario with Mean Matching-Graph Distance of 150\label{fig:jsjy_a}}
    {\includegraphics[width=.8\textwidth]{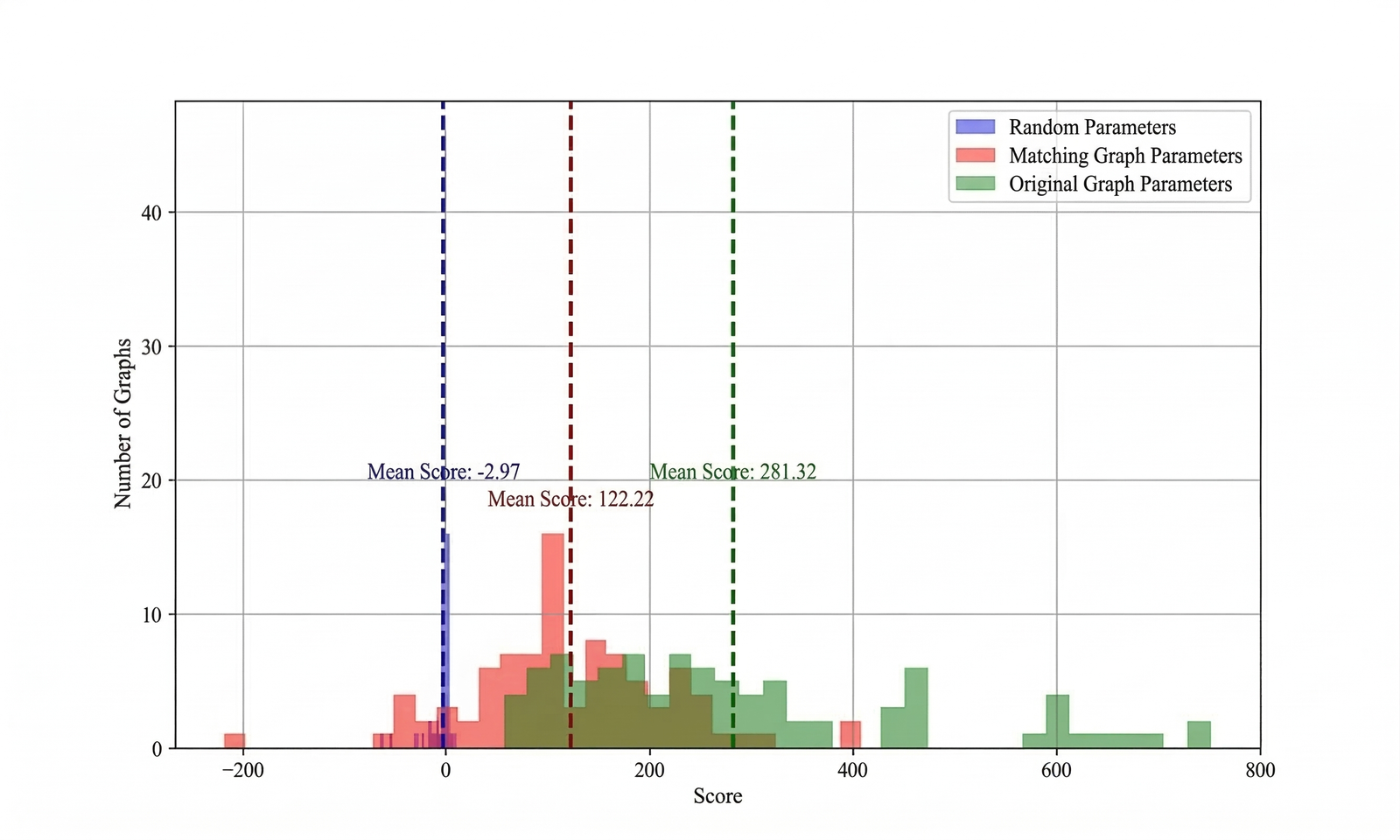}}
    \subcaptionbox{Scenario with Mean Matching-Graph Distance of 100\label{fig:jsjy_b}}
    {\includegraphics[width=.8\textwidth]{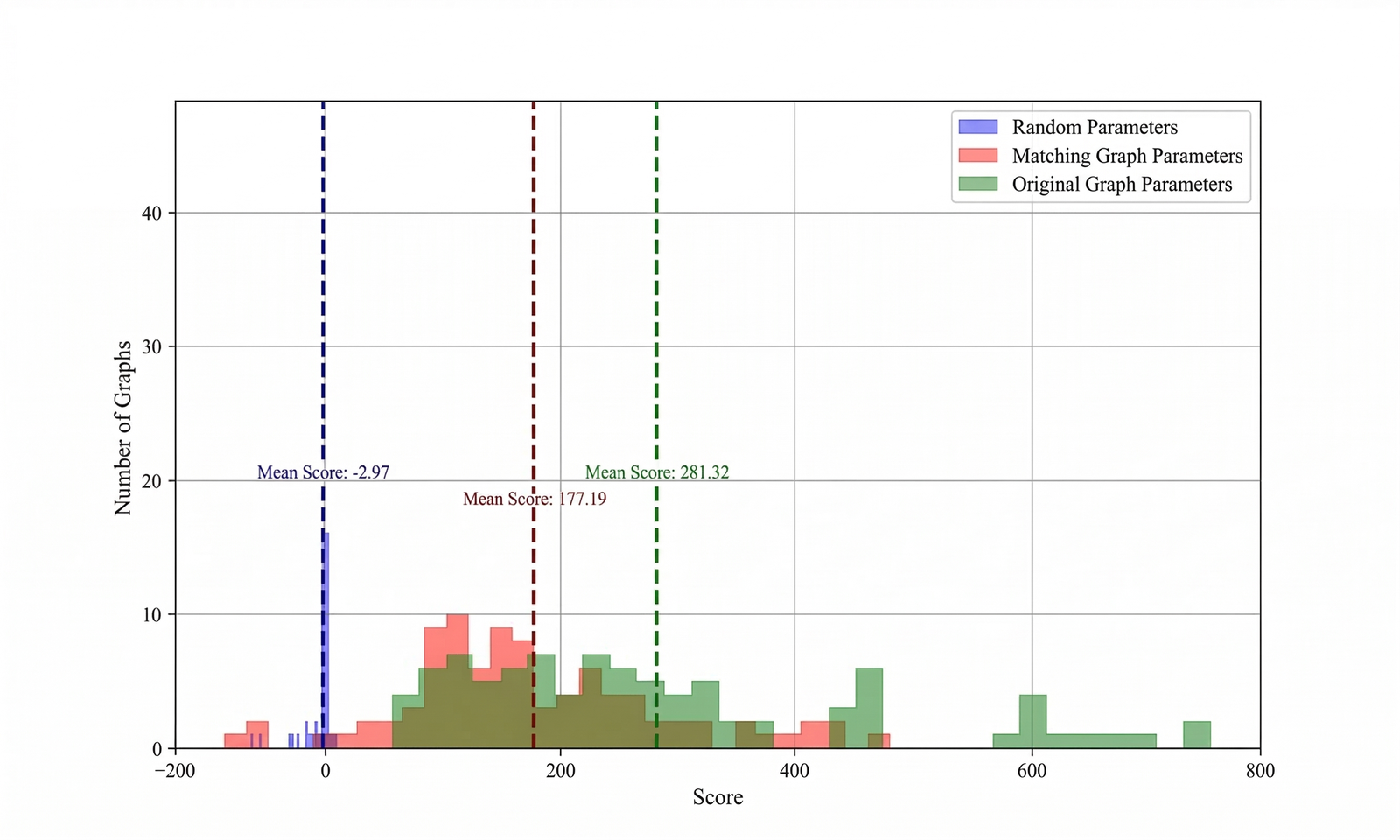}}
    \caption{Scores Obtained by Transferring Random Parameters, Matching Graph Parameters, and Original Graph Parameters}\label{fig:jsjy}
\end{figure}

\begin{figure} 
    \centering 
    \includegraphics[width=0.9\textwidth]{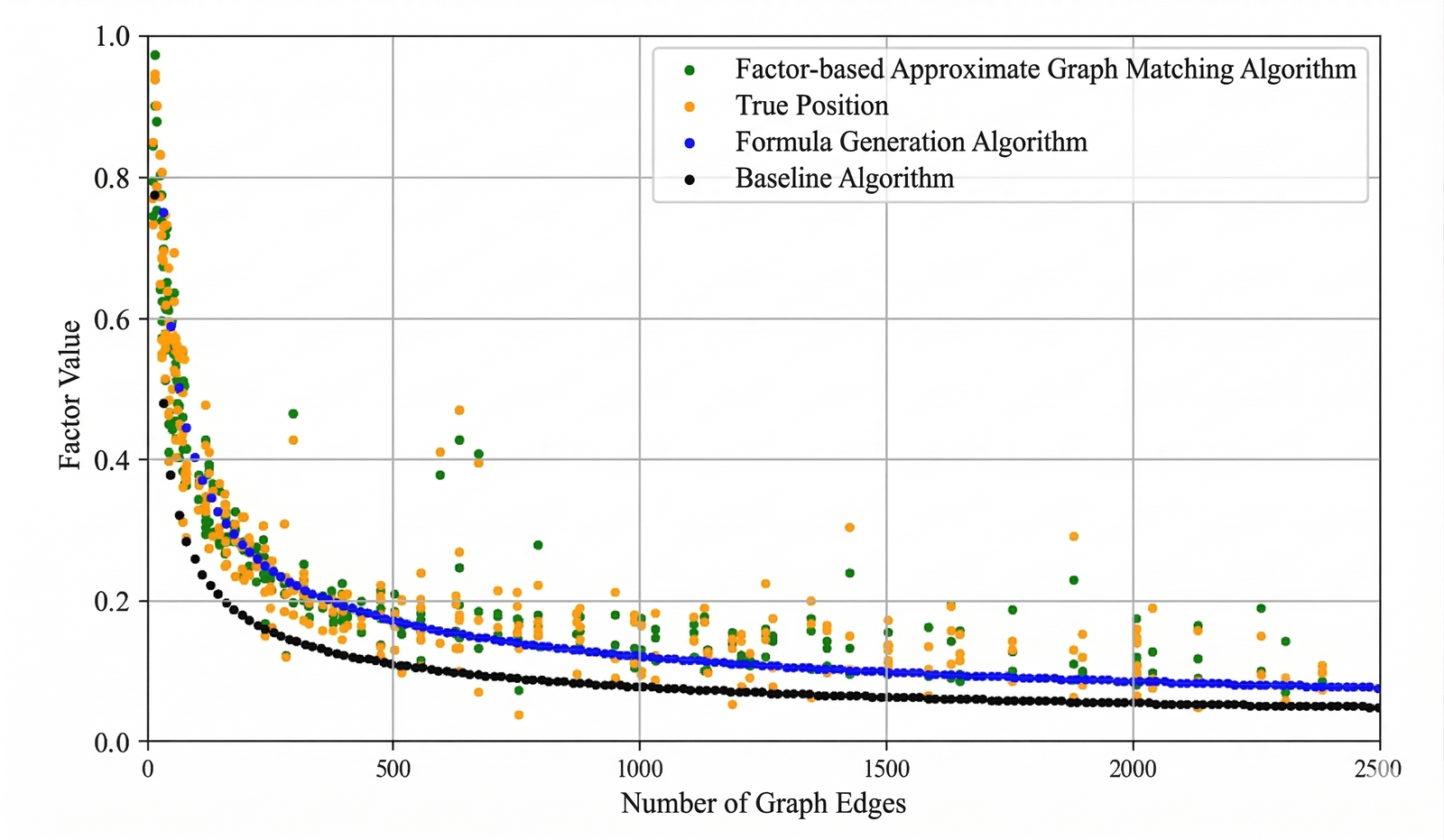}
    \caption{Factor Distributions of Different Algorithms}
    \label{fig:factor}
\end{figure}

(1) Transferability of Parameters from Matched Approximate Graphs
\\ \indent (a) Scores of Transferred Parameters from Homologous Graphs
\\ \indent In this experiment, 8 graphs are randomly selected from the local dataset as the Original Graphs. Using the approximate graph matching function in the accelerator's parameter generation module, homologous graphs are found for them, that is, graphs with distance 0, referred to as Homologous Graphs. The optimized parameters of the Homologous Graphs are then applied to the corresponding Original Graphs, and the score comparison shown in Table~\ref{tab:yzxdb} is obtained.
\\ \indent In the table, the second row gives the exact parameter scores obtained by running multiple rounds of QAOA optimization on each Original Graph, the third row gives the scores obtained by applying the parameters of the algorithm-matched Homologous Graphs, and the fourth row gives the score differences.
It can be seen that the scores of the Homologous Graph parameters are very close to the exact parameter scores of the Original Graphs, and the total difference across the 8 graphs is only 53.868. The relative difference is only 2.67\%. It is even observed that, for Graph 7, the exact parameter score obtained by running the standard QAOA algorithm is slightly worse than the score obtained by transplanting the parameters from its matched Homologous Graph.
This verifies both the transferability of Homologous Graph parameters that coincide with the coordinates of the Original Graphs and the effectiveness of the proposed approximate graph matching algorithm.
\begin{table}[htbp]
    \centering
    \caption{Scores of Original Graph Parameters and Transferred Parameters from Algorithm-Matched Homologous Graphs}
      \resizebox{\textwidth}{!}{%
      \begin{tabular}{|c|c|c|c|c|c|c|c|c|c|}
      \toprule
            & Graph 1    & Graph 2    & Graph 3    & Graph 4    & Graph 5    & Graph 6    & Graph 7    & Graph 8    & Total \\
      \midrule
      Original Graph    & 66.044 & 291.591 & 219.183 & 96.473 & 233.221 & 245.909 & 262.813 & 599.561 & 2014.795 \\
      \midrule
      Homologous Graph   & 62.969 & 278.279 & 212.509 & 96.302 & 228.928 & 236.931 & 263.087 & 581.922 & 1960.927 \\
      \midrule
      Difference    & 3.075 & 13.312 & 6.674 & 0.171 & 4.293 & 8.978 & -0.274 & 17.639 & 53.868 \\
      \bottomrule
      \end{tabular}%
      }
    \label{tab:yzxdb}%
  \end{table}%
\\ \indent (b) Scores and Statistical p-values of Random Parameters, Matching Graph Parameters, and Original Graph Parameters
\\ \indent Based on the previous experiment, a more rigorous hypothesis test can be conducted to further verify parameter transferability and the effectiveness of the proposed algorithm. In the dataset, 90 graphs are randomly selected as Original Graphs. Then, graphs with close distances are matched to them as approximate graphs, without requiring them to be homologous. The scores obtained by transplanting Random Parameters and the parameters of the matched approximate graphs, referred to as Matching Graph Parameters, to the Original Graphs are then compared.
\\ \indent Figure~\ref{fig:jsjy} shows the comparison histograms of the scores of Random Parameters, Matching Graph Parameters, and Original Graph Parameters. The blue histogram shows the scores under Random Parameters, the red histogram shows the scores obtained by transplanting Matching Graph Parameters, and the green histogram shows the exact parameter scores of the Original Graphs.
From Figure~\ref{fig:jsjy_a}, it can be seen that the mean score of the Random Parameters is -2.97, and almost all of them are concentrated near 0, with almost no overlap with the Original Graph Parameters, shown in green.
By contrast, the red region representing the Matching Graph Parameters, when the mean distance of the Matching Graphs is 150, covers most of the green region. The mean score of the Matching Graph Parameters is 122.22, which indicates that, compared with Random Parameters, Matching Graph Parameters can greatly improve parameter quality.
At this point, the statistical p-value between the Random Parameters and the Matching Graph Parameters is $3.95\times 10^{-24}$, which also shows that the Matching Graph Parameters perform significantly better than the Random Parameters.
\\ \indent In addition, as the mean distance of the Matching Graphs decreases, as shown in Figure~\ref{fig:jsjy_b}, when the mean distance is reduced to 100, the red dashed line representing the mean score of the Matching Graph Parameters shifts rightward to 177.19.
It can also be noted that the variance of the red region becomes smaller. This indicates that, as the matched graphs become closer to the Original Graphs, the quality of the Matching Graph Parameters can be further improved. At this point, the statistical p-value between the Random Parameters and the Matching Graph Parameters is $2.14\times 10^{-34}$, which further confirms this conclusion.
\\ \indent (2) Effectiveness of the Factor-based and Formula Generation Algorithms
\\ \indent The default code of this MindSpore Quantum Computing Hackathon implements the algorithm of Ref. \cite{Sureshbabu_2024}, which is the so-called Baseline Algorithm.
As mentioned in the previous chapters, both the Factor-based Approximate Graph Matching algorithm and the Formula Generation algorithm are improvements over the Baseline Algorithm.
The reason for adjusting the factor $\arctan \frac{1}{\sqrt{D-1}}$ in the Baseline Algorithm is that empirical observation shows that the factor used by the Baseline Algorithm introduces bias in parameter prediction.
\\ \indent Figure~\ref{fig:factor} shows the distribution of factor values for the graphs in the dataset. Since the factor values are mainly related to the number of graph edges, and the factors in both the Baseline Algorithm and the Formula Generation algorithm are designed with degree as the parameter, the distribution is plotted with the Number of Graph Edges on the horizontal axis.
In the figure, the black points represent the factor values generated by the Baseline Algorithm, the blue points represent those generated by the proposed Formula Generation algorithm, the green points represent those generated by the Factor-based Approximate Graph Matching algorithm, and the orange points represent the exact factor positions, that is, the True Position marked in the figure.
\\ \indent First, Figure~\ref{fig:factor} shows that the factor value decreases as the number of graph edges increases, with a rapid decline at the beginning and a gradually slowing trend afterward. This is basically consistent with the functional shape of the Baseline Algorithm factor formula $\arctan \frac{1}{\sqrt{D-1}}$;
\\ \indent Second, from the distribution of the True Position, it can be seen that the factor values generated by the Baseline Algorithm deviate from the true values. The black points lie clearly below most of the orange points, which means that the parameter values generated by the Baseline Algorithm are generally too small.
This is precisely why the Formula Generation algorithm of Stone-in-Waiting adjusts the coefficient of the Baseline Algorithm. For the dataset of this competition, a coefficient of 1.56 is multiplied onto all factors generated by the Baseline Algorithm. Of course, this coefficient is only an empirical value for this competition and can be adjusted according to different
scenarios. As can be seen in the figure, the adjusted factor values, represented by the blue points of the Formula Generation algorithm, are closer to the orange points. Therefore, compared with the Baseline Algorithm, the Formula Generation algorithm ultimately produces better parameter quality.
It can also be observed that the proposed Formula Generation algorithm is closer to the True Position when the number of graph edges is relatively small. In the figure, the blue points are more consistent with the orange points when the number of edges is small, while they are slightly lower when the number of edges is larger. This is because the local dataset provided in this competition mainly contains graphs with fewer than 1000 edges,
so when setting the coefficient, greater weight is given to the true factor positions of graphs with fewer than 1000 edges;
\\ \indent Third, the figure shows that the Factor-based Approximate Graph Matching algorithm has a good ability to fit the True Position. The green points are basically distributed around the orange points, and some of them almost overlap. This also indirectly verifies that the algorithm is capable of generating high-quality parameters;
\\ \indent Finally, it should be mentioned that, although the Formula Generation algorithm produces parameter quality that is only moderate compared with the Factor-based and Parameter-based algorithms, as can be seen from the relatively limited coverage of the orange-point region by the blue points in the figure,
it nevertheless has two advantages. On the one hand, the Formula Generation algorithm has a clear advantage over the Baseline Algorithm. On the other hand, it does not require any pre-built database of computed parameters or factors and can directly output parameter values on demand. This gives the algorithm good generalization performance and universality.
In Stone-in-Waiting, this algorithm is used as a fallback when the other algorithms perform poorly. In other words, the parameter quality produced by the Formula Generation algorithm guarantees the lower bound of the output parameter quality of the accelerator.
\subsection{Experiments on System Performance}
(1) Overall Score and Analysis
\\ \indent For convenience, the score of the Baseline Algorithm on the 90 competition graphs in the \texttt{data} directory, as evaluated by the official \texttt{score.py} program, is called the Baseline Local Score. The score of the Baseline Algorithm on the online \texttt{data/\_hidden} directory is called the Baseline Online Score. Their sum, namely the score reported by the competition scoring system, is called the Baseline Total Score.
Tests show that the Baseline Algorithm achieves a Local Score of 16526.80, an Online Score of 8473.11, and a Total Score of 24999.91.
\\ \indent Up to now, Stone-in-Waiting version 0.0.1v has achieved a Local Score of 25318.87, an Online Score of 9727.63, and a Total Score of 35046.50. The Total Score is 40.19\% higher than the Baseline Total Score.
It is worth noting that this score is not the final result. In fact, from the difference in score improvement between the Baseline Algorithm and Stone-in-Waiting on the local and online datasets, it can also be seen that, compared with the online dataset, Stone-in-Waiting produces better parameter quality on the local dataset.
This is because the current database of computed parameters maintained by the accelerator has much denser coverage over the local dataset than over the online dataset.
Since Stone-in-Waiting is designed with functions such as Continuous Parameter Search and Reverse Parameter Updating, the database of computed parameters will gradually expand as the cloud accelerator continues to run, and the quality of the output parameters for the online dataset will also be further improved.

\begin{figure} 
    \centering 
    \includegraphics[width=0.9\textwidth]{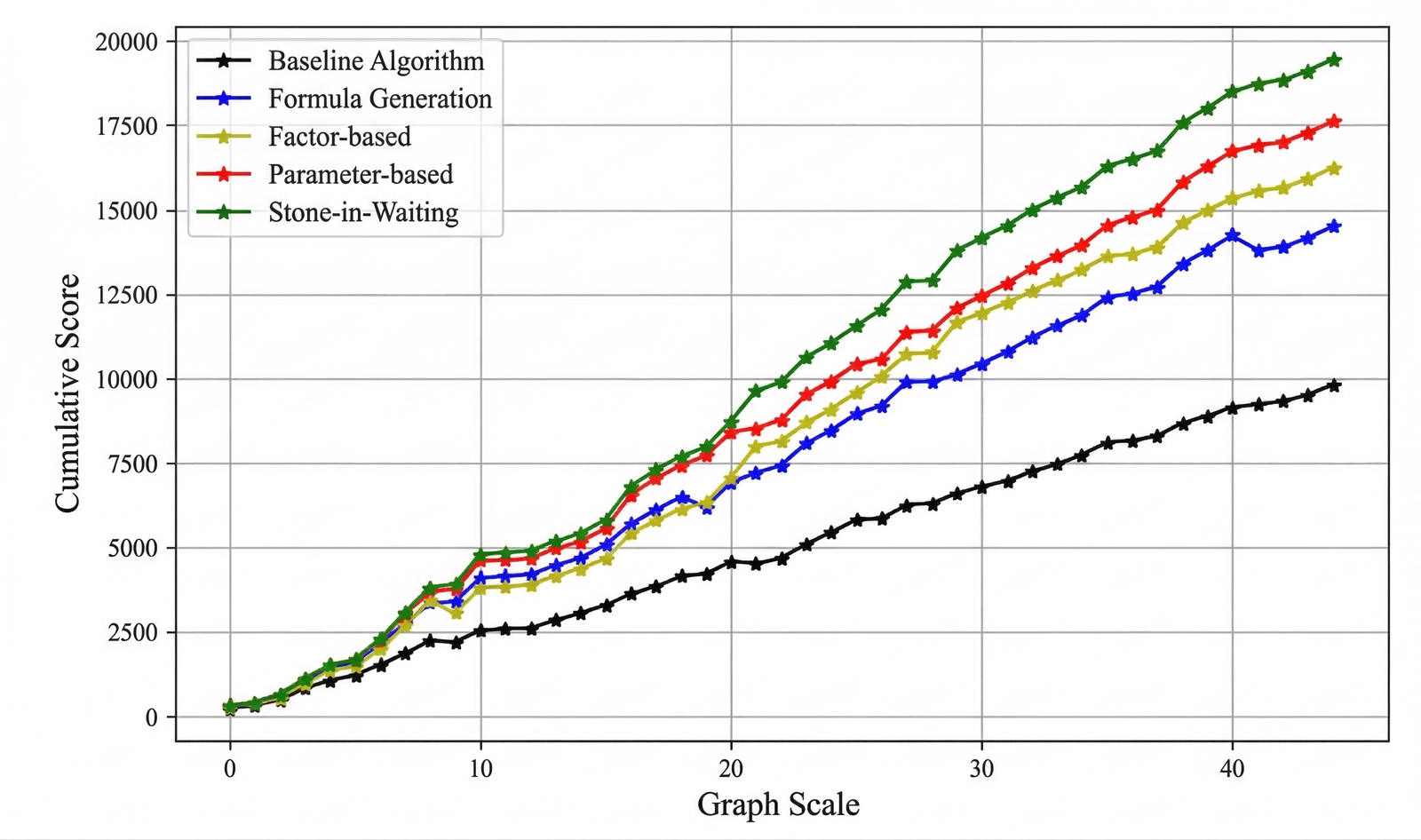}
    \caption{Algorithm Performance Comparison ($S_p=1000$, $S_o=400$)}
    \label{fig:sfxndb}
\end{figure}

\begin{figure}
\centering
\subcaptionbox{$S_p=500$, $S_o=200$\label{fig:sfxndb_dcj_a}}
{\includegraphics[width=.45\textwidth]{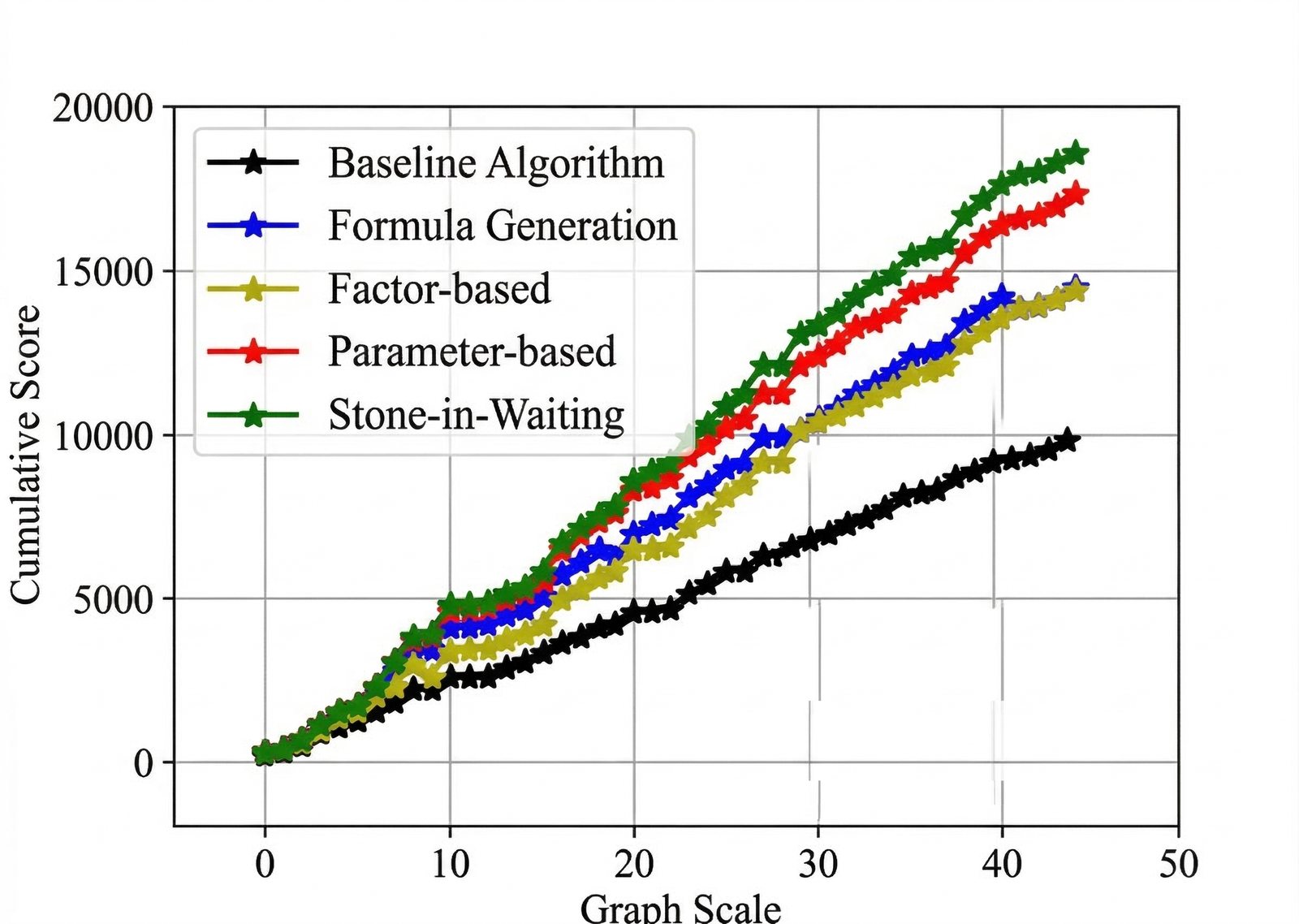}}
\subcaptionbox{$S_p=250$, $S_o=100$\label{fig:sfxndb_dcj_b}}
{\includegraphics[width=.45\textwidth]{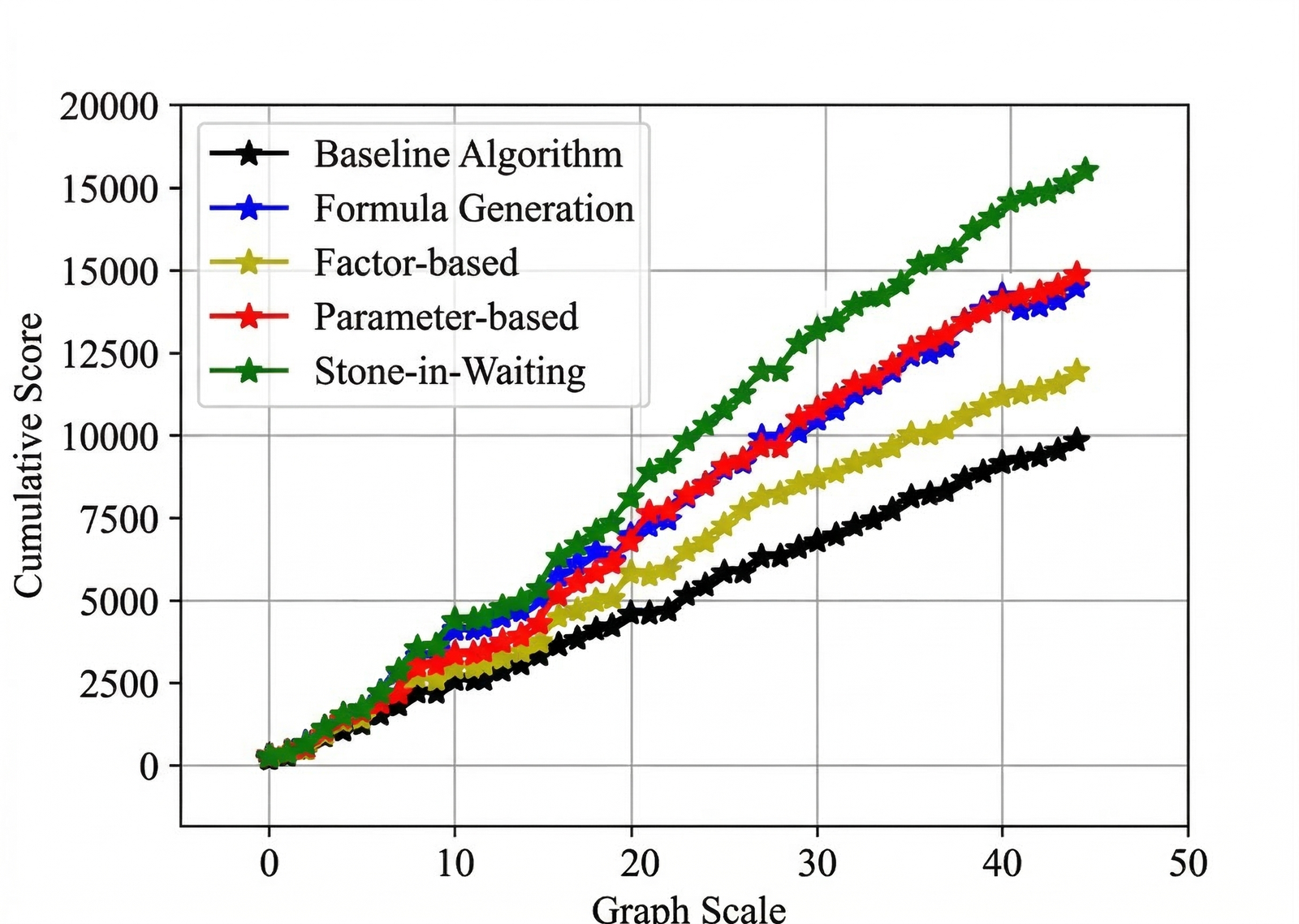}}
\subcaptionbox{$S_p=125$, $S_o=50$\label{fig:sfxndb_dcj_c}}
{\includegraphics[width=.45\textwidth]{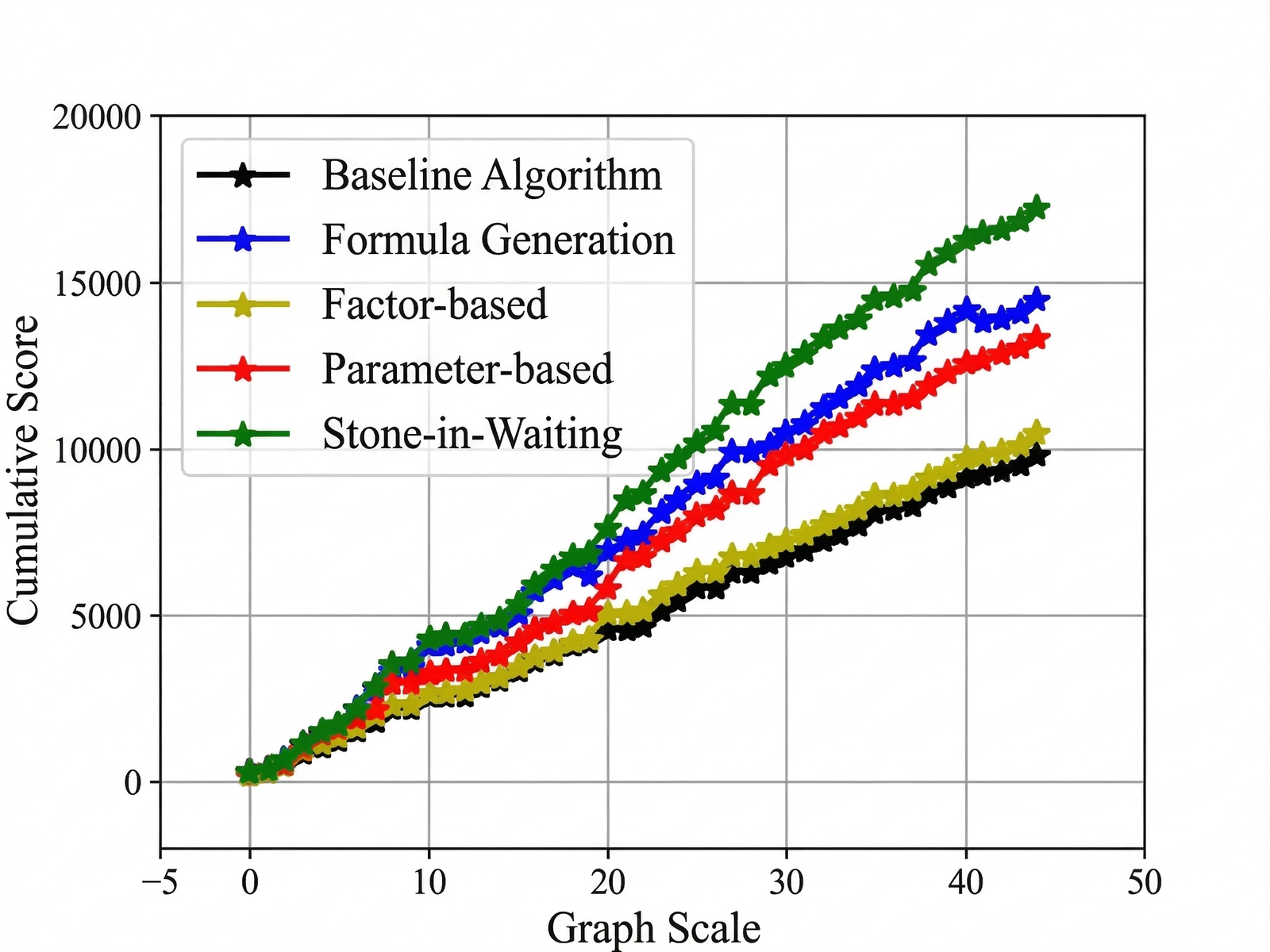}}
\subcaptionbox{$S_p=50$, $S_o=50$\label{fig:sfxndb_dcj_d}}
{\includegraphics[width=.45\textwidth]{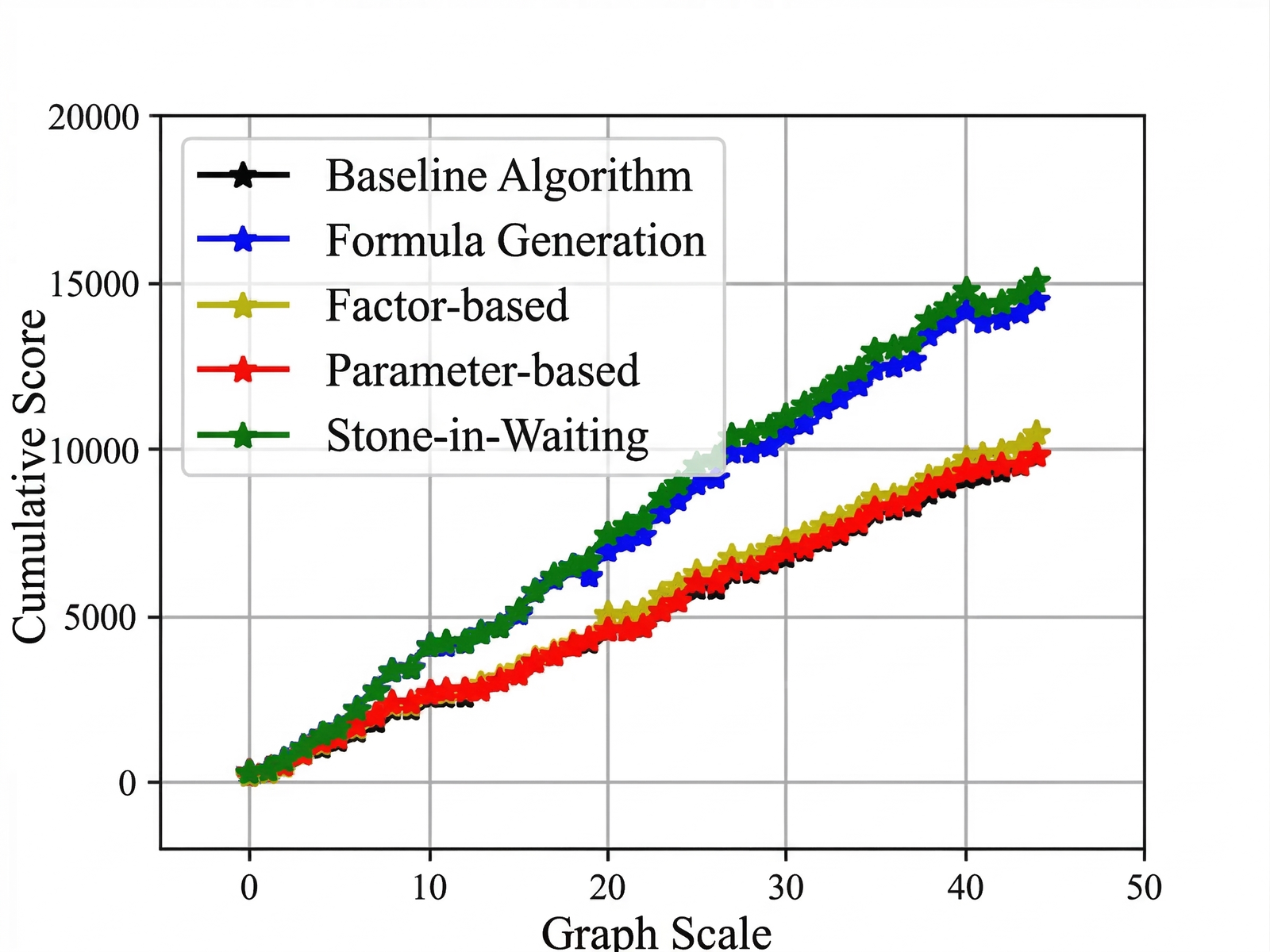}}
\caption{Algorithm Performance Comparison under Different Scenarios}\label{fig:sfxndb_dcj}
\end{figure}

(2) Algorithm Performance and Comparison
\\ \indent For this test, 45 graphs are randomly selected as Original Graphs. Parameters are then generated with the Baseline Algorithm, Stone-in-Waiting, and the three sub-algorithms in its parameter generation module. These sub-algorithms are the Parameter-based method, the Factor-based method, and the Formula Generation method. The resulting scores are accumulated for comparison.
At this point, the size of the database of computed parameters is set to 1000, denoted by $S_p$, and the size of the database of computed factors is set to 400, denoted by $S_o$.
\\ \indent Figure~\ref{fig:sfxndb} shows the performance of the different algorithms. In the figure, the black curve represents the Cumulative Score produced by the Baseline Algorithm, the green curve represents the Cumulative Score produced by Stone-in-Waiting, and the red, yellow, and blue curves represent the Cumulative Scores produced by the Parameter-based method, the Factor-based method, and the Formula Generation method, respectively.
It can be seen that, first, the Cumulative Scores of Stone-in-Waiting and all the proposed sub-algorithms are higher than those of the Baseline Algorithm, and the gap increases with Graph Scale. Although only 45 random graphs are shown here, similar performance advantages can also be observed on the local dataset and even on the full dataset;
Second, since Stone-in-Waiting selects the best output among the different sub-algorithms and uses it as the final output algorithm of the accelerator, its performance naturally remains above that of all the individual algorithms;
Finally, under the conditions of this experiment, that is, $S_p=1000$ and $S_o=400$, among the three proposed sub-algorithms, the Parameter-based method performs slightly better, the Factor-based method is second, and the Formula Generation method is slightly worse than the first two.
\\ \indent It is worth noting that the curves of the three proposed sub-algorithms in Figure~\ref{fig:sfxndb} overlap to some extent during their rise. This is not accidental. In fact, under different scenarios, the sub-algorithms may each have their own advantages and disadvantages, which is precisely the reason and role of the integrated best-selection mechanism in Stone-in-Waiting.
\\ \indent The overlap in algorithm performance can be highlighted by testing several typical scenarios. Figure~\ref{fig:sfxndb_dcj} includes the performance of the algorithms under four scenarios. Figure~\ref{fig:sfxndb_dcj_a} is obtained by halving the sizes of the databases of computed parameters and computed factors relative to the setting of Figure~\ref{fig:sfxndb}.
That is, the scenario changes from that of Figure~\ref{fig:sfxndb}, with $S_p=1000$ and $S_o=400$, to that of Figure~\ref{fig:sfxndb_dcj_a}, with $S_p=500$ and $S_o=200$.
It can be seen from the figure that, compared with Figure~\ref{fig:sfxndb}, the Parameter-based Approximate Graph Matching algorithm, the Factor-based Approximate Graph Matching algorithm, and Stone-in-Waiting as a whole all show some performance degradation because of the reduced coverage of the database of computed parameters.
Among them, the decline of the Factor-based algorithm is more pronounced than that of the Parameter-based algorithm. The yellow and blue curves almost overlap, and the performance of the Factor-based algorithm drops to a level close to that of the Formula Generation algorithm. This is because, although the size of the database of computed parameters is reduced, its density still has some redundancy, and the current scenario can still cover the region of the Original Graphs relatively well.
\\ \indent When the database density is further reduced to $S_p=250$ and $S_o=100$, as shown in Figure~\ref{fig:sfxndb_dcj_b}, the red curve representing the Parameter-based algorithm also drops to a position close to the blue curve representing the Formula Generation algorithm. This indicates that the further halving of the size of the database of computed parameters has a relatively large impact on algorithm performance.
At this point, an interesting phenomenon is that, although both the Parameter-based and Factor-based algorithms decline significantly, the performance of Stone-in-Waiting is almost unaffected. Note that the heights of the green curves in Figure~\ref{fig:sfxndb_dcj_a} and Figure~\ref{fig:sfxndb_dcj_b} hardly change. The reason for this phenomenon is that Stone-in-Waiting selects
the best output among the sub-algorithms, so although individual sub-algorithms may exhibit some local performance degradation, their combined effect is still sufficient to sustain the overall performance of the accelerator in the current scenario.
\\ \indent As the database sizes are halved again, in the scenario of Figure~\ref{fig:sfxndb_dcj_c}, where $S_p=125$ and $S_o=50$, the performance of both the Factor-based and Parameter-based algorithms declines further. Both fall below the Formula Generation algorithm, and the Factor-based algorithm becomes very close to the Baseline Algorithm.
At this point, the performance of Stone-in-Waiting is also slightly affected and declines accordingly.
\\ \indent When the database of computed parameters becomes even sparser and decreases to $S_p=50$, as shown in Figure~\ref{fig:sfxndb_dcj_d}, the performance of the Parameter-based algorithm also drops to a level close to that of the Baseline Algorithm. At this point, the performance of Stone-in-Waiting is very close to that of the Formula Generation algorithm, and the green and blue curves almost overlap. When $S_p=0$ and $S_o=0$, they would overlap completely.
This is precisely because the Formula Generation algorithm has good universality, and its performance does not vary with the size of the database of computed parameters, so it plays a strong fallback role for Stone-in-Waiting.
\\ \indent In general, the above experiments demonstrate the performance of the parameter generation sub-algorithms under different scenarios and test their scope of applicability as well as their respective strengths and weaknesses. These experiments further prove the necessity and effectiveness of integrating the sub-algorithms in Stone-in-Waiting.
\section{Conclusion and Future Work}
This paper proposes and designs a cloud-based accelerator named Stone-in-Waiting to rapidly obtain high-quality initial parameters for the Quantum Approximate Optimization Algorithm. Building on the most advanced current theories and methods for parameter determination, Stone-in-Waiting internally designs and implements four self-developed algorithms. By selecting the best output among them, the system can generate parameters that match given graph data quickly and accurately.
Experimental results show that, compared with the Baseline Algorithm, the accelerator improves the parameter score by 40.19\%. In addition, Stone-in-Waiting provides both a web interface and an API, offering flexible and convenient support for users to test and develop related experiments and applications.
\\ \indent More importantly, based on the principle of approximate-graph parameter transferability established in frontier literature and further validated in this paper, Stone-in-Waiting attempts to build a new quantum-computing paradigm through functions such as continuous parameter search and reverse parameter updating.
That is, reusable resources are recovered from the results of parameter optimization for approximate optimization algorithms.
This process is analogous to the energy-recovery system of hybrid vehicles. If properly utilized and scaled into a broader practice or even an industry convention, it may significantly improve the efficiency of the Quantum Approximate Optimization Algorithm and the utilization of quantum computing resources in the future.

\section*{Acknowledgement}
This work was supported by the CPS-Yangtze Delta Region Industrial Innovation Center of Quantum and Information Technology-MindSpore Quantum Open Fund.